\documentstyle[preprint,aps,epsf]{revtex}

\begin{document}

\title{Statistical Fluctuations of  Electromagnetic Transition Intensities
in $pf$-Shell Nuclei}

\author{A. Hamoudi$^1$, R.G. Nazmitdinov$^1$ and Y. Alhassid$^2$}
\address{$^1$Bogoliubov Laboratory of Theoretical Physics,
Joint Institute for Nuclear Research, 141980 Dubna, Russia \\
$^2$Center for Theoretical Physics,
Sloane Physics Laboratory, Yale University, New Haven,
Connecticut 06520,  USA}
\date{\today}
\maketitle

\begin{abstract}

We study the fluctuation properties of  $E2$ and $M1$
transition intensities  among $T=0,1$ states of $A = 60$ nuclei  in the
framework
of the interacting shell model, using a realistic effective interaction for
 $pf$-shell nuclei with a $^{56}Ni$  core.
It is found that the $B(E2)$ distributions are well described by the Gaussian
orthogonal ensemble of random matrices (Porter-Thomas distribution)
independently of the isobaric quantum number $T_z$.    However, the
 statistics of the $B(M1)$ transitions is sensitive to   $T_z$:   $T_z=1$ nuclei
 exhibit a Porter-Thomas distribution, while
a significant deviation from the GOE statistics is
observed for self-conjugate nuclei ($T_z=0$).

\end{abstract}

\draft

\narrowtext

 Random matrix theory (RMT) \cite{Meh} was originally introduced to
explain the statistical fluctuations of  neutron resonances observed in
 compound nuclei \cite{Por}.
  The theory assumes that the nuclear Hamiltonian belongs to an ensemble of
 random matrices that are consistent
 with the fundamental symmetries of the system. In particular, since the
nuclear interaction preserves time-reversal symmetry, the relevant ensemble
 is the Gaussian orthogonal ensemble (GOE).  Bohigas  {\it et al} \cite{Boh}
conjectured that  RMT describes the statistical fluctuations
 of a quantum system whose associated classical dynamics is chaotic.
More recently, a proof for this conjecture was proposed by mapping the problem
 on the  supersymmetric non-linear  sigma model \cite{AASA96}.
RMT has become a universal tool for analyzing statistical fluctuations
in chaotic systems \cite{Br,Guh}.

 The  chaotic nature of the single-particle dynamics  in the nucleus can be
 studied in the framework of the mean-field approximation.  The interplay
between shell
 structure and  fluctuations in the single-particle spectrum can be understood
in terms of the classical dynamics of the nucleons in the corresponding
  deformed mean-field potential \cite{Ar,H94}.
However, the residual nuclear interaction mixes different mean-field
configurations and plays a crucial role  in determining the nature of the
 statistical fluctuations of the many-particle spectrum and wavefunctions.
 The statistics  of the
 low-lying collective part of the nuclear spectrum has been studied in the
framework of the interacting boson model, in which
 the nuclear fermionic space is mapped onto a much smaller
space of  bosonic degrees of freedom \cite{Al1,Al2}.
 Because of the relatively
small number of degrees of freedom in this model, it was also possible to
relate the statistics to the underlying nature of  the mean-field collective
dynamics.   Since at higher excitations additional degrees of freedom (such
as broken pairs) become important \cite{Al4},
 it is necessary  to analyze  the effect of interactions on
the statistics in much larger model spaces.
The interacting shell model offers an attractive framework for such studies
where realistic effective interactions are available (in finite model spaces),
 and the basis states are labeled by  exact quantum numbers of angular
 momentum,  isospin and parity \cite{Zel}.

 RMT makes definite predictions for the statistical  fluctuations of  the
 eigenfunctions as well as the spectrum.  The  electromagnetic
 transition intensities in a nucleus
are observables that are sensitive to the wavefunctions, and
 the investigation
of their statistical  distributions should complement the more common
 spectral analysis \cite{Al1,Al2}.
$B(M1)$ and $B(E2)$ transitions were recently analyzed
in $^{22}Na$ \cite{m1} in the framework of the shell model
 and found to follow  the Porter-Thomas distribution
\cite{PT}, in agreement with RMT and consistent with the previous finding of
a Gaussian distribution for the eigenvector components \cite{v1}-\cite{v4}.
In addition, the distributions of  $E2$ and $M1$  matrix elements did not
 show any sensitivity to  spin and isospin \cite{m1}.

 Most  studies of statistical  fluctuations in the shell model have been
restricted
to lighter nuclei  ($A \alt 40$) where complete $0 \hbar \omega$  calculations
 are feasible (e.g. $sd$-shell nuclei). It is of interest to investigate
 how the statistics evolves with increasing  mass number.
In the present paper we study the fluctuation properties
of  electromagnetic transition intensities  in nuclei with $A \sim 60$.
 We find that the $B(E2)$ distributions are Porter-Thomas, but that the
$ M1$ statistics is sensitive to isospin. In particular the $B(M1)$
distributions
 in self-conjugate nuclei show a significant deviation from Porter-Thomas.
The calculations are performed in the $pf$ shell with $^{56}$Ni as
 a core, i.e. we  assume a fully occupied $f_{7/2}$ orbit  and consider
  all possible many-nucleon configurations defined by the
$0f_{5/2}$, $1p_{3/2}$ and $1p_{1/2}$ orbitals. The effective interaction
is chosen to be the isospin-conserving F5P interaction \cite{in1}.
This interaction is successful in describing the mass range $A \sim 57-68$.
The calculations were performed with
 the shell model program OXBASH \cite{2}.

 Denoting by $B(\overline{\omega}L; i \to f)$ the reduced transition probability
 from an initial state $|i \rangle$ to a final state $|f \rangle$, with
 $\bar{\omega}$ indicating the electric ($E$) or magnetic ($M$)
character of the transition, and $2^{L}$ the multipolarity, we have \cite{Bru}
\begin{equation}
\label{bel}
B(\overline{\omega}L; J_i T_i T_z \rightarrow J_f T_f T_z)=
\frac{|\delta_{T_i T_f}M_{is}(\overline{\omega}L)
-  (T_i T_z 10|T_f T_z ) M_{iv}(\overline{\omega}L)|^2} {(2J_i+1)(2T_i+1)} \;.
\end{equation}
Here  $M_{is}(\overline{\omega}L)$
and $M_{iv}(\overline{\omega}L)$ are the triply reduced matrix elements
 for the isoscalar and isovector components of  the transition operator,
 respectively.  Note that these matrix elements depend on $J_i, T_i$
 and $J_f, T_f$ but not on $T_z$.
 For $\Delta T=0$ transitions ($T_i=T_f =T$) the
isospin Clebsch-Gordan coefficient in Eq. (\ref{bel})  is simply given by
\begin{equation}
(T T_z 10|T T_z )={T_z}/\sqrt{T(T+1)} \;.
\end{equation}
 It follows that  the isovector component  in Eq. (\ref{bel}) is absent
 for self-conjugate  (i.e. $T_z=0$) nuclei. Thus the statistics of the
 isoscalar component
of an electromagnetic transition operator  can be inferred
directly from  $T_z=0$ nuclei.   Consequently, we can test
the sensitivity of the statistics to
the isovector and the isoscalar contributions.

To study the fluctuation properties of
the transition rates, it is necessary to divide out any secular variation
 of the average strength function versus the initial and final
energies.  We calculate  an average transition
strength at an initial energy $E$ and
final energy $E^\prime$ from \cite{Al3,Al2}
\begin{eqnarray}
\label{av1}
\langle B(E,E^\prime) \rangle=\frac{\sum_{i,f}B(\overline{\omega}L; i
\rightarrow f) e^{-(E-E_i)^2/2\gamma^2 }
e^{-(E^\prime-E_f)^2/2\gamma^2}}{\sum_{i,f}e^{-(E-E_i)^2/2\gamma^2}
e^{-(E^\prime-E_f)^2/2\gamma^2}} \;.
\end{eqnarray}
 For fixed values of  the initial ($J_i^{\pi},T$) and final ($J_f^{\pi},T$)
 spin, isospin and parity, we calculate from Eq. (\ref{bel})  the intensities
$B(\overline{\omega}L; i\rightarrow f)$. All transitions  of a given operator
(e.g. $M1$ or $E2$) between the initial and final states of the  given
spin, parity and isospin classes have been included in the statistics.
We remark that the energy levels used in (\ref{av1}) are the unfolded energy
levels\cite{BG}, characterized by a constant mean spacing.
The value of $\gamma$ has been chosen to be large enough
to minimize effects arising from
the local fluctuations in the transition strength but not
so large  as  to wash away the secular energy variation of the average
intensity.  In the present calculations we used  $\gamma=4$.
 We renormalized
the actual intensities by dividing out their smooth part
\begin{equation}\label{av2}
y_{fi}= {B(\overline{\omega}L;  J_i T T_z  \to J_f T T_z) \over
\langle B(E,E^\prime \rangle} \;,
\end{equation}
and  histogrammed them into  bins equally spaced in
$\log y$.  In RMT there is a large number of weak transitions, and we use a
 logarithmic scale to display small values of  $y$ over several orders of
magnitude.  For each transition operator and  classes of  initial and final
states  we fit to the calculated
distribution a  $\chi^2$ distribution in $\nu$ degrees of freedom \cite{Lev}
\begin{equation}\label{fit}
P_{\nu}(y)=(\nu /2<y>)^{\nu /2}y^{\nu /2 - 1}e^{-\nu y/2<y>}/\Gamma(\nu/2) \;.
\end{equation}
For $\nu=1$ this distribution reduces to the Porter-Thomas distribution. In
systems
with mixed classical dynamics, it is found that $\nu$ decreases monotonically
 from $1$ as the system makes a transition from chaotic to regular motion
\cite{Al3}.

\begin{figure}

\epsfxsize=13 cm
\centerline{\epsffile{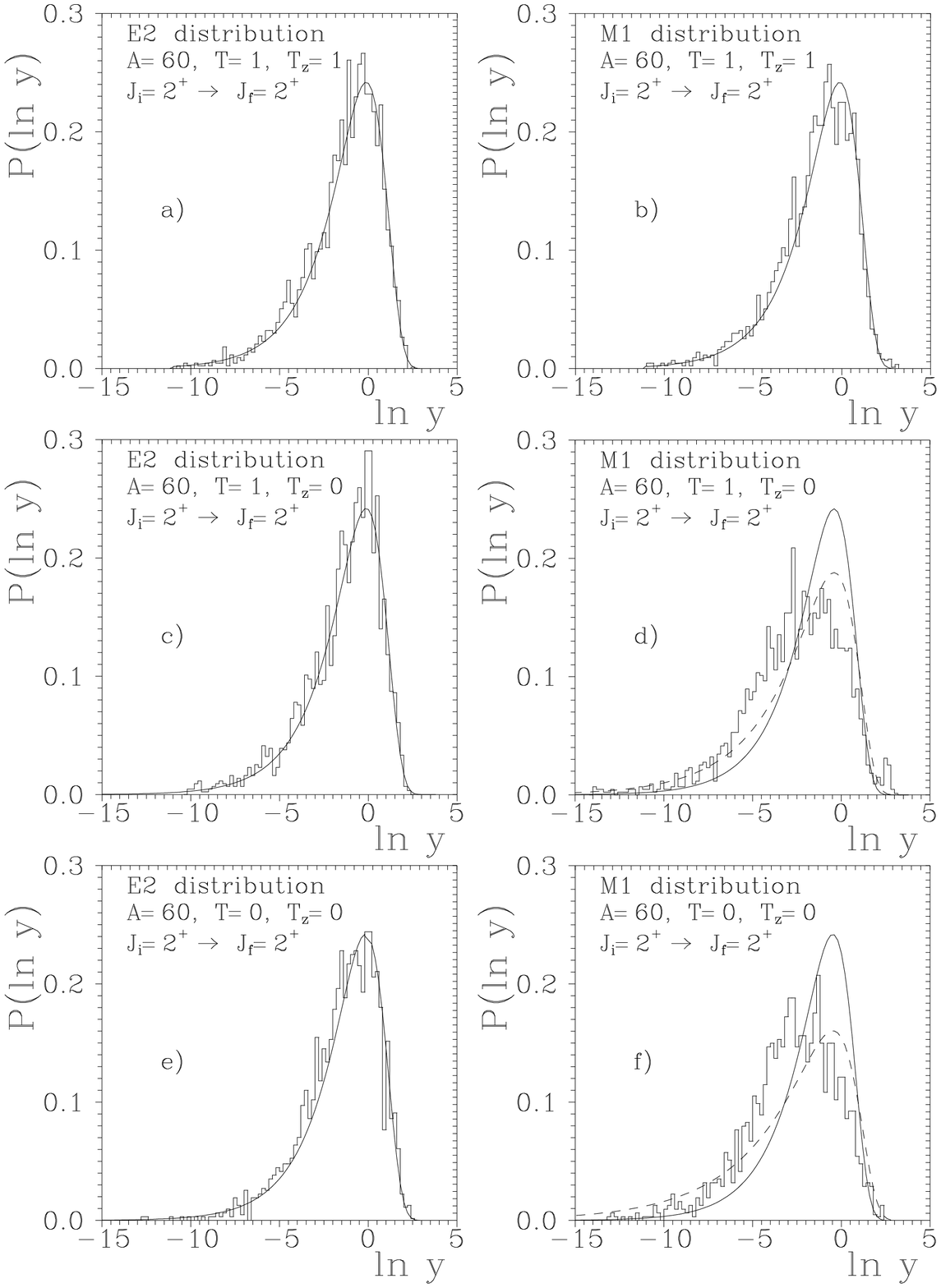}}

\vspace{2 mm}

\caption
{ The  $B(E2)$  and $B(M1)$ intensity distributions (histograms) for the
 $2^{+}, T \rightarrow 2^{+}, T$ transitions  in $A= 60$ nuclei:
(a,b) $2^{+}, 1 \rightarrow 2^{+}, 1$ transitions in $^{60}$Co  ($T_z=1$);
(c,d) $2^{+}, 1 \rightarrow 2^{+}, 1$ transitions in $^{60}$Zn  ($T_z=0$);
(e,f) $2^{+}, 0 \rightarrow 2^{+}, 0$ transitions in $^{60}$Zn  ($T_z=0$).
   The solid  lines describe the Porter-Thomas distribution
(Eq. (\protect\ref{fit}) with $\nu=1$). The dashed lines in d)  and f) are
Eq. (\protect\ref{fit}) with   $\nu=0.64$  and $\nu=0.54 $, respectively .}
\label{fig1}

\newpage

\epsfxsize=14 cm
\centerline{\epsffile{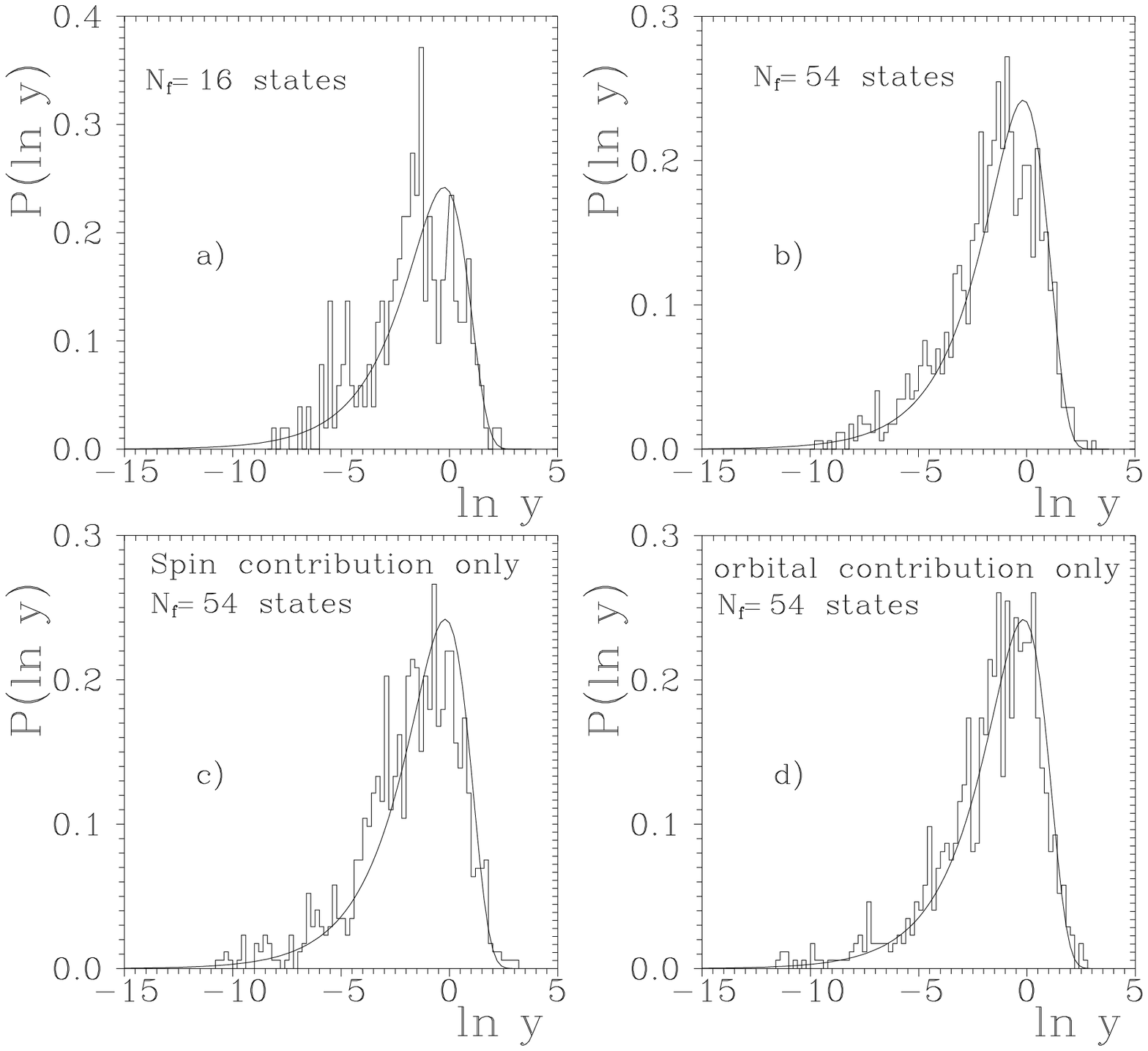}}

\vspace{2 mm}

\caption
{ a,b) Distributions of $M1$ intensities for the
 $0^{+}, 1\rightarrow 1^{+}, 1$  transitions in $^{60}$Co  ($T_z=1$)
with different number of final states included in the statistics (see text).
c,d) The distributions of the spin and orbital contributions to  $M1$,
for the same transitions as in (b).
The solid lines are the Porter-Thomas distributions.}
\label{fig2}

\newpage

\epsfxsize=14 cm
\centerline{\epsffile{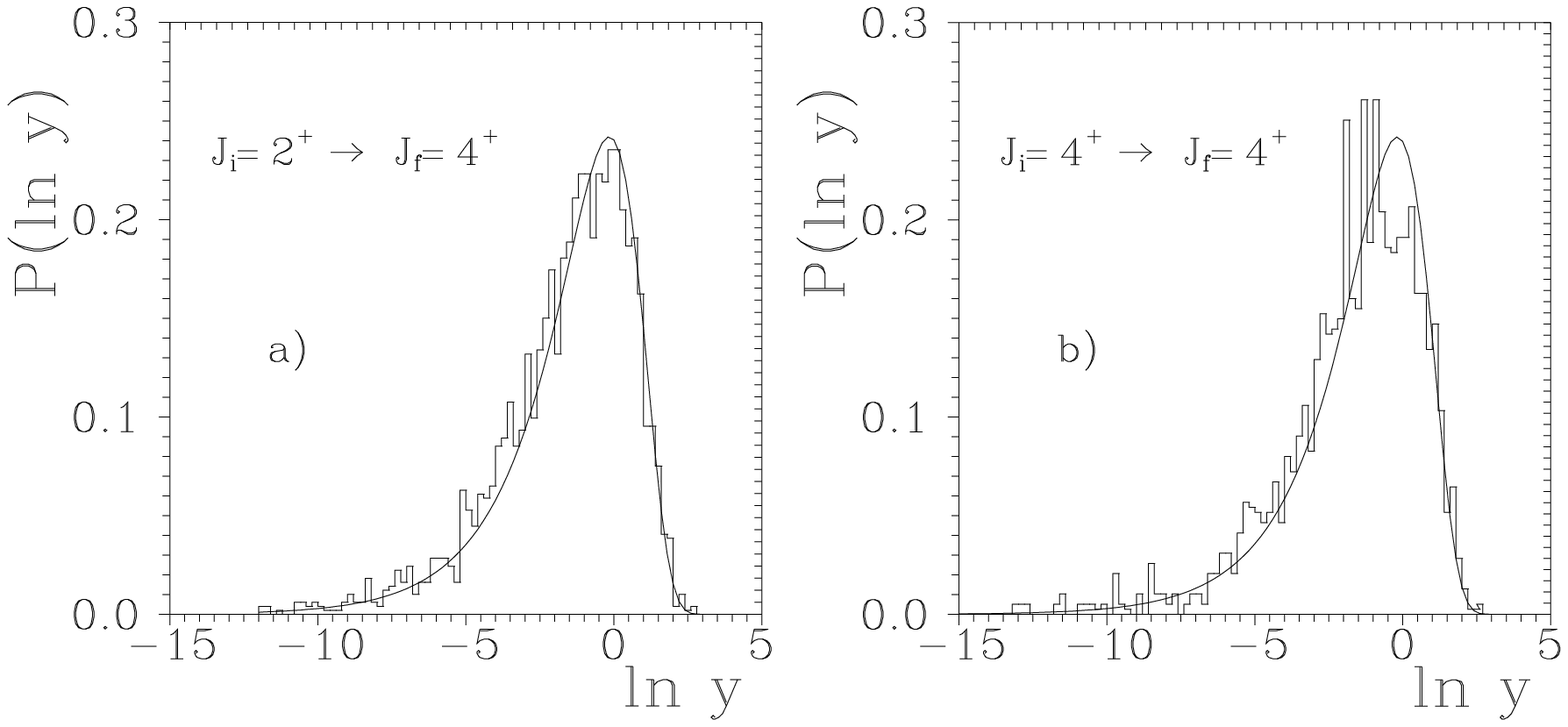}}

\vspace{2 mm}

\caption
{The $B(E2)$ intensity distributions for the transitions:
a) $2^{+}, T=0 \rightarrow 4^{+}, T=0$; b) $4^{+}, T=0\rightarrow 4^{+},
 T=0$ in $^{60}$Zn.}
\label{fig3}

\end{figure}

We first examine  all $E2$ and $M1$   $2^{+}, T\rightarrow 2^{+}, T$
transitions ($T=0$ or $T=1$) in $A=60$ nuclei.   $T=0$ states exist only in
$T_z=0$ nuclei, i.e.
$^{60}$Zn in our case. However, the $T=1$ states form isobaric multiplets in
$^{60}$Co, $^{60}$Zn and $^{60}$Cu.  In the latter case we studied the
statistics in both $T_z=0$ and $T_z=1$ nuclei.  Because of the vanishing
of the isovector Clebsch-Gordan coefficient  in Eq. (\ref{bel}), the
transitions in
$^{60}$Zn ($T_z=0$) are purely isoscalar.  For each transition operator we
sampled  $56^2=3136$ and $66^2=4356$ intensities
for $T_z=0$ and $T_z=1$, respectively.  The calculated  distributions
 (histograms) of the $B(E2)$  (left panels) and $B(M1)$ (right panels)
 $2^{+}, T \rightarrow 2^{+}, T$ transitions are shown in Fig. \ref{fig1}, and
compared with the Porter-Thomas distribution (Eq. (\ref{fit}) with $\nu=1$).

 The distributions of transitions within the $T=1$ states are shown
 in the  top ($T_z=1$) and middle ($T_z=0$) panels.  Since our interaction
  is isospin invariant, the
energy levels of an isobaric multiplet are degenerate, and the spectral
statistics of  $T=1$ states must be the same in both $T_z=0$ and $T_z=1$
nuclei.  We find  that the level spacing and $\Delta_3$ statistics are in
agreement with GOE.
However, while the $B(E2)$ distributions are all in agreement with the
GOE prediction,  the $M1$ distributions show strong sensitivity to $T_z$.
For $T_z=1$ the distribution of the  $M1$ intensities is Porter-Thomas, but in
 self-conjugate nuclei we find that the $M1$ distribution deviates significantly
from the
GOE limit.  The dashed line in Fig. \ref{fig1}d shows a $\chi^2$
 distribution (\ref{fit}) with $\nu=0.64$. Although it does not fit well the
 calculated distribution, the smaller $\nu$ is consistent with the
 observed larger number of weak transitions as
 compared with a Porter-Thomas distribution.   A similar deviation is
 observed in the $M1$ distribution for
the $2^{+}, T=0 \rightarrow 2^{+}, T=0$ transitions (see Fig. \ref{fig1}f),
while the $B(E2)$ probabilities are distributed according to
Porter-Thomas (Fig. \ref{fig1}e). The deviation from the GOE occurs
for the $M1$ transitions in self-conjugate nuclei (e.g. $^{60}$Zn), where
 the matrix elements are purely isoscalar.
In  $T_z=1$ nuclei  both isoscalar and isovector components contribute
 to the $M1$ transitions. However, since the isoscalar $M1$ matrix elements
 are much weaker than the isovector M1 \cite{Bru}, the latter dominate and the
 distributions  are restored to their GOE form.

 We also investigated the statistics  of the $0^+,  T=1\to 1^+,  T=1$
transitions and its dependence on the number of  final states.
In the top panels of Fig. \ref{fig2} we compare the  $M1$ distribution
$(T_z=1)$ between all $16$ $0^+,  T=1$ initial states and the lowest
 $16$  $1^+,  T=1$ final states  with the corresponding distribution when all
$54$ final states are taken into account.  Both seem to give similar
distributions, but the statistics is  poorer in the first case.

 The $M1$ strength function is composed of a spin contribution
 and an orbital contribution, each dominating in a different energy region.
   We find  that  the orbital part is
 in close agreement with the Porter-Thomas statistics  while the dominating
spin part shows  a slight deviation (see Fig. \ref{fig2}c,d).

Finally, we tested the sensitivity of the statistics to the initial and final
spin. In Fig. \ref{fig3} we show the $B(E2)$
 distributions for the
$2^{+} \rightarrow 4^{+}$ and $4^{+}\rightarrow 4^{+}$  transitions
 between states with $T=0$.  The $B(E2)$ distribution for the
 $2^{+} \rightarrow 4^{+}$ transitions is well described by Porter-Thomas,
as in the $2^{+} \rightarrow 2^{+}$ case
of Fig. \ref{fig1}e  (in both cases $J_i=2^+$).  However,
the $4^{+}\rightarrow 4^{+}$
 transitions do show a small deviation from the Porter-Thomas limit.
This suggests that the class of $4^+$ states is slightly less chaotic.

In conclusion, we have studied the distributions of  $B(M1)$  and of  $B(E2)$
transition strengths in $pf$-shell nuclei with  $A \sim 60$. While most
 of the calculated distributions are in close agreement with the
 Porter-Thomas distribution
 predicted by the  GOE, we find that the $M1$ transitions in
self-conjugate nuclei ($T_z =0$) deviate significantly  from the
RMT limit. For these nuclei the $M1$ transitions are purely isoscalar and
 relatively weak compared with transitions  in $T_z=1$  nuclei, which are
 dominated by the isovector component of $M1$.

The authors are grateful for useful discussions with F. Dittes.
We thank V. Ponomarev for assistance
during the preparation of the manuscript. This work was supported
in part by the Department of Energy grant No. DE-FG-0291-ER-40608.

\newpage

\end{document}